\documentclass[11pt]{article}

\usepackage{jheppub}
\usepackage[T1]{fontenc}
\usepackage[utf8]{inputenc}
\usepackage{amsfonts,amsmath,amssymb,accents,mathrsfs}
\usepackage[mathscr]{euscript}
\usepackage[retainorgcmds]{IEEEtrantools}
\usepackage{multirow}
\usepackage{multicol}
\usepackage{tikz}
\usepackage{graphicx,color}
\usepackage{booktabs}
\usepackage{placeins}
\usepackage{mathtools}
\usepackage{calc}
\usepackage[greek,english]{babel}
\usepackage{youngtab}
\usepackage[sort&compress]{natbib}
\usepackage[colorlinks=true,
			linktocpage=true,
            linkcolor=blue,
            citecolor=blue,
            urlcolor=blue,
            ]{hyperref}

\renewcommand\a {{\alpha}}
\renewcommand\b {{\beta}}
\newcommand\g {{\gamma}}
\renewcommand\d {{\delta}}

\newcommand\s {{\sigma}}

\renewcommand\t {{\tau}}
\newcommand\e {{\epsilon}}
\renewcommand\th {{\theta}}
\renewcommand\L {{\Lambda}}
\renewcommand\l {{\lambda}}

\newcommand\N{{\mathcal{N}}}

\newcommand\V{{\mathcal{V}}}
\newcommand\W{{\mathcal{W}}}
\newcommand\U{{\mathcal{U}}}
\newcommand\eW{{\mathscr{W}}}
\newcommand\eU{{\mathscr{U}}}

\newcommand\tth {{\tilde{\theta}}}
\newcommand\tD {{\tilde{\D}}}
\newcommand\tX {{\tilde{X}}}
\newcommand\tT {{\tilde{T}}}

\newcommand\D{{\rm D}}

\newcommand\pa{{\partial}}
\def\un{\underline}

\def\fracm#1#2{\hbox{\large{${\frac{{#1}}{{#2}}}$}}}

\def\bea{\begin{IEEEeqnarray*}}
\def\eea{\end{IEEEeqnarray*}}
\def\be{\begin{eqnarray}}
\def\ee{\end{eqnarray}}
\def\n{\IEEEyesnumber}
\def\sn{\IEEEyessubnumber}

\makeatletter
\renewcommand\section{\@startsection{section}{1}{\z@}
              {3ex plus-1ex minus-.2ex}{1pt plus1pt}
              {\large\sf\bfseries\boldmath}}
\renewcommand{\subsection}{\@startsection{subsection}{2}{\z@}
              {1.5ex plus-1ex minus-.2ex}{0.01pt plus1pt}{\sf\slshape}}
\renewcommand{\subsubsection}{\@startsection{subsubsection}{3}{\z@}
              {1.5ex plus-1ex minus-.2ex}{0.01pt plus0.2pt}{\sf\boldmath}}
\renewcommand{\paragraph}{\@startsection{paragraph}{4}{\z@}
              {.75ex \@plus.5ex \@minus.2ex}{-2mm}{\sf\bfseries\boldmath}}
\makeatother

\title{\boldmath Nonlinear $\N=2$ Supersymmetry and\\ 3D Supersymetric Born-Infeld Theory}

\author{Yangrui Hu and}
\author{Konstantinos Koutrolikos}

\affiliation{Brown Theoretical Physics Center \\
Department of Physics, Brown University}

\emailAdd{yangrui\_hu@alumni.brown.edu}
\emailAdd{konstantinos\_koutrolikos@brown.edu}

\abstract{
D$p$-branes acquire effective nonlinear descriptions whose bosonic part is related to the Born-Infeld
action. This nonlinearity has been proven to be a consequence of the partial $\N=2\to\N=1$ supersymmetry
breaking, originating from the solitonic nature of the branes.  In this work, we focus on the effective
descriptions of D2-branes.  Using the Goldstone multiplet interpretation of the action and the method of
nilpotent $\N=2$ superfields, we construct the 3D, $\N=1$ superspace effective action which makes the first
supersymmetry manifest and realizes the second, spontaneously broken, supersymmetry nonlinearly. We show that
there are two such supersymmetric extensions of the 3D Born-Infeld action which correspond to the dynamics of
the 3D Maxwell-Goldstone multiplet and the 3D projection of the Tensor-Goldstone multiplet respectively.
Moreover, we demonstrate that these results are derived by applying the constrained superfield approach on the
$\N=2, D=3$ vector and chiral multiplets after expanding them around a nontrivial vacuum.  We find that these
two descriptions are related by a duality transformation which results in the inversion of a dimensionless
parameter. For both descriptions we derive the explicit bosonic and fermionic parts of the 3D super
Born-Infeld action.  Finally, consider the deformation of the Maxwell-Goldstone superspace action by the
characteristic Chern-Simons-like, gauge invariant, mass term.
}

\dedicated{\textgreek{για τον Πάνο}}
\preprint{\phantom{Brown}}
\begin{document}

\maketitle
% \flushbottom

%\tableofcontents

% our new things:
% [26,27]: they started from 4D to 3D projection and coset method, the extra one translation sym -> central charge (we don't have that); we consider 3D N=2 to N=1
% component field result
% nilpotent condition

% unique or not, explore the mass term deformation: one parameter \xi deformation (or you can absorb it into mass, then mass becomes the one free parameter)

% duality -> inversion relation of lambda : extend BI to include dilation...fluxes U(1) -> SL(2,R), for SL(2,R), there is inversion transf (strong-weak duality, so inversion is interesting)

% outlook: TTbar discussion for future work

% modifications:
% modify the paragraph in introduction
% summary section, add outlook,
% ackownledgement

\section{Introduction}
\label{sec:intro}

The deep connection between partial supersymmetry breaking and nonlinear realizations of extended
supersymmetries has been studied extensively
\cite{Rocek:1978nb,Deser:1980ck,Cecotti:1986gb,Hughes:1986dn,Hughes:1986fa,Antoniadis:1995vb,Bagger:1996wp,Bagger:1997pi,Rocek:1997hi,Antoniadis:2008uk,Kuzenko:2015rfx}.
One of the most transparent demonstrations of this connection is the effective description of D$p$-branes
which are solitonic solutions of string theory \cite{Dai:1989ua,Leigh:1989jq}. The introduction of a boundary
to the world-sheet theory breaks some of the spatial translational symmetries and half of the supersymmetries
\cite{Polchinski:1995mt} which implies that D-branes correspond to BPS saturated soliton states.  These
solutions acquire a lower-dimensional effective description where the bosonic part is given by the DBI action
for the corresponding collective coordinates and the world volume gauge field
\cite{Leigh:1989jq,Townsend:1995af,Schmidhuber:1996fy,Gibbons:1997xz,deAlwis:1996ze}. This action is closely
related to the nonlinear Born-Infeld type of action of open string theory
\cite{Fradkin:1985qd,Abouelsaood:1986gd,Bergshoeff:1987at,Metsaev:1987qp} because the DBI action for all
$p$-branes ($p<9$) always corresponds to the dimensional reduction of the ten-dimensional (D9-brane) BI action
\cite{Gibbons:1997xz}.

The existence of these effective nonlinear descriptions is no accident and follows directly from the partial
supersymmetry breaking $\N=2\to\N=1$. In general, the memory of a spontaneously broken symmetry does not fade
completely and is captured through nonlinear realizations of that symmetry.  In this fashion, the effective
description of D$p$-branes must be consistent with the linear representations of the surviving supersymmetry
as well as with the nonlinear realizations of the second, broken supersymmetry.  Such nonlinear realizations
of supersymmetry are defined via nonlinear constraints which generate precisely the Born-Infeld type actions
that appear in the effective actions of D-branes.

For the D3-brane, the manifestly $\N=1$ supersymmetric effective action was first found in
\cite{Deser:1980ck,Cecotti:1986gb}. However, as it was shown, demanding the surviving supersymmetry to be
manifest is not enough to uniquely determine the superspace action. Later in \cite{Bagger:1996wp} the 4D,
$\N=1$ supersymmetric Born-Infeld (BI) action was derived based on the understanding that it will correspond
to the dynamics of the massless Goldstone supermultiplet associated with the partial breaking of the second
supersymmetry. In addition to the $\N=1$ superspace description, demanding the invariance of the theory under
the second non-linearly realized supersymmetry fixed all previous ambiguities and determined the superspace
Lagrangian.

A general method for constructing nonlinear realizations of a symmetry is to start with a linear
representations of that symmetry and impose consistent nonlinear constraints. For the case of supersymmetry
this led to the constrained superfield approach \cite{Rocek:1978nb} which can be applied universally to find
nonlinear actions.  In \cite{Rocek:1997hi} using the constrained superfield approach it was shown that the
results of \cite{Bagger:1996wp} directly follow from the $\N=2$ free Maxwell theory by (\emph{i}) expanding
around a non-trivial vacuum that breaks the second supersymmetry and (\emph{ii}) imposing the standard
nonlinear --nilpotency-- constraint which reduces the field content of the $\N=1$ theory to the
Maxwell-Goldstone multiplet by relating the $\N=1$ chiral and vector multiplet components.

Interestingly, it has been shown \cite{Antoniadis:2008uk,Kuzenko:2009ym} that demanding invariance under the
nonlinear supersymmetry does not uniquely determine the superspace Lagrangian as expected. There exist a
consistent deformation of the theory by adding a Fayet-Iliopoulos (FI) term to the superspace action of
\cite{Bagger:1996wp,Rocek:1997hi}. The FI term respects the invariance under the second non-linearly realized
supersymmetry and is $\N=1$ manifest. However, its effect is to give a nonzero VEV to the auxiliary component
of the vector multiplet and therefore it spontaneously breaks the first supersymmetry. Nevertheless, the
deformed theory still enjoys an unbroken $\N=1$ symmetry which corresponds to a different choice of an $\N=1$
subsector of the $\N=2$ theory than the one that is made manifest.

It is often common practice, once an action is found, to use dualities ---symmetries between the equations of
motion and their Bianchi identities--- to map it to a dual action with equivalent on-shell dynamics. The
duality transformations can be applied to a large class of (non-Gaussian) actions which depend only on the
field strength of a p-form algebraically. For the special case of $p=(d-2)/2$ it may happen that the dual
action is the same functional\footnote{Up to additional transformations of background fields and parameters.}
of the dual field strength as the original action for the original field strength. The BI action and its 4D
supersymmetric extension \cite{Bagger:1996wp,Rocek:1997hi} falls in this category of self-dual theories with
the duality transformation being the standard electric-magnetic duality.  By turning on the background dilaton
and axion fields\footnote{One should also turn on the background two-form $C_{2}$ and four-form $C_4$ of the
R-R sector and the NS-NS two form $B_2$. However, they are not necessary for observing the presence of the
SL(2,R) duality group.}, the U(1) duality rotation is extended to SL(2,R) which is the self-duality group of
the effective D3-branes \cite{Tseytlin:1996it}. The supersymmetric theory with the FI deformation can be
understood as a special case of the supersymmetric extension of this generalized BI action
\cite{Kuzenko:2009ym}.

Similar to the special role of the D3-brane in IIB theory, D2-brane plays a significant role in IIA theory.
What makes them particularly interesting is that they can be interpreted as the (M-theory) eleven-dimensional
supermembrane \cite{Bergshoeff:1987cm,Townsend:1995af,Schmidhuber:1996fy}. One method of deriving the 3D
$\N=1$ supersymmetric extension of the effective Born-Infeld action associated with the D2-brane is to
dimensionally reduce the 4D results of \cite{Bagger:1996wp} down to 3D in a way that breaks half the
supercharges.  This is the path followed in \cite{Ivanov:1999fwa,Ivanov:2001gd}\footnote{We are grateful to
E.~Ivanov for bringing these papers to our attention.} via the use of the \emph{coset} approach
\cite{Callan:1969sn,Volkov:1973vd} which provides a systematic method of constructing nonlinear realizations
and studying properties of Goldstone fields.  In this case, the authors considered appropriate factorizations
of elements of the coset space $\{\N$=$1$, $D$=$4$ ~\textit{Super-Poincar\'{e}}\}/SO(1,2) which reflect the
spontaneously breaking of one translational symmetry, one supersymmetry and the remaining rotations in
$SO(1,3)/SO(1,2)$. As demonstrated in \cite{Ivanov:1999fwa}, the Goldstone superfields associated to the
broken supersymmetry and rotation generators are not independent and can be expressed covariantly in terms of
the Goldstone superfield associated to the broken translational symmetry.  Although the various transformation
properties and dynamical equations of motion were derived, it is emphasized that this methodology does not
permit the construction of a non-trivial superspace action for the independent Goldstone superfield. For this
task the methodology of \cite{Bagger:1997pi} was followed. In \cite{Ivanov:2001gd}, the above results are
understood as the outcome of applying the general method of deriving nonlinear realizations of supersymmetry
from appropriate linear ones presented in \cite{Ivanov:1978mx,Ivanov:1989bh,Delduc:1999in,Ivanov:2000nk}, on a
particular deformation of the 3D $\N=2$ Maxwell multiplet proposed in \cite{Zupnik:1999qy}.

In this work, we derive the 3D $\N$=$1$ superspace action corresponding to the 3D BI theory ---associated to
the effective description of the D2-brane--- by considering appropriate and manifestly $\N$=$2$, $D$=$3$
theories which  are decomposed to $\N$=$1$, $D$=$3$, massless Goldstone multiplets via the constraint
superfield approach.  We identify two such $\N=2$ multiplets, the vector and the chiral which give rise to the
3D, $\N=1$ Maxwell-Goldstone and the 3D, $\N=1$ Tensor-Goldstone multiplets respectively.  For both
descriptions, we derive explicitly the bosonic and fermionic parts of the supersymmetric spacetime actions and
we show that they are related by a duality transformation which generates an inversion
$\lambda~\to~\tilde{\lambda}=1/\lambda$ of a characteristic dimensionless parameter $\lambda$.  Furthermore,
we demonstrate that the ---unique to 3D--- gauge invariant, Chern-Simons-like mass term of the
Maxwell-Goldstone action generates one parameter family deformations which respect the non-linearly realized
supersymmetry. However this deformation explicitly breaks the first, linear, supersymmetry.

The layout of this paper is the following. In section \ref{sec:MG} we construct the 3D Maxwell-Goldstone
multiplet by following the arguments of Bagger-Galperin \cite{Bagger:1996wp} and reproduce the results of
\cite{Ivanov:1999fwa}. Subsequently, we show that this Goldstone multiplet is the result of a nilpotent,
$\N=2$ superfield generated by the expansion of the $\N=2$ Maxwell multiplet around a nontrivial vacuum.
Starting from the effective, superspace action we derive the explicit spacetime effective action. As expected,
the bosonic part gives the 3D Born-Infeld action. The fermionic part of the Lagrangian has a more complicated
structure where polynomial terms are weighted by nonpolynomial factors which correspond to derivatives of the
3D Cecotti-Ferrara function.  In section \ref{sec:deformation}, we consider the Chern-Simons mass term of the
vector multiplet as a generator of a one-parameter deformation of the Maxwell-Goldstone action and show that
such a deformation is consistent with the second, nonlinear supersymmetry but violates the first
supersymmetry.  In section \ref{sec:TG}, we use the  3D Tensor-Goldstone multiplet in order to construct the
$\N=1$ supersymmetric extension of the 3D Born-Infeld action. We show that this Goldstone multiplet
corresponds to the application of the nilpotent superfield approach on the  3D, $\N=2$ chiral multiplet and
similar to section \ref{sec:MG} we extract the explicit bosonic and fermionic parts of the spacetime
Lagrangian.  Finally, in section \ref{sec:duality} we show that the 3D $\N=1$ Maxwell-Goldstone and
Tensor-Goldstone multiplets as well as their corresponding superspace actions, map to each other under duality
transformations. This duality generates an inversion of the dimensionless parameter $\lambda$ constructed out
of the characteristic dimensionful parameters of the theories. We conclude with the summary and two
appendices.

\section{3D, ${\cal N}=1$ Maxwell-Goldstone Multiplet}
\label{sec:MG}

\subsection{Review of 3D, ${\cal N}=1$ Maxwell Multiplet}
In three dimensions, the $\N=1$ vector multiplet is described by the spinorial superfield strength $W_\a$
constrained by the Bianchi identity:
\begin{equation}\label{MBI}
    \D^\a\,W_{\a} ~=~ 0 ~~\Rightarrow~~ \D^2\,W_\a~=~i\pa_\a{}^\b\, W_\b ~~.
\end{equation}
This constraint can be solved by expressing the superfield strength $W_\a$ in terms of an unconstrained
---\emph{prepotential}--- spinorial superfield $\Gamma_\a$:
\be\label{W}
W_\a ~=~ \frac12 \,\D^{\b}\,\D_{\a}\,\Gamma_{\b}~~.
\ee
The prepotential $\Gamma_\a$ is not uniquely defined and gives rise to an equivalence class $[~\Gamma_\a~]$,
where the equivalence relation corresponds to the following gauge transformation
\begin{equation}
     \d\Gamma_\a ~=~ \D_\a\,K ~~
    \label{equ:gauge-Gamma}
\end{equation}
where $K$ is an arbitrary scalar superfield.
The superspace action takes the form
\be\label{Svm}
S~=~\frac{1}{g^2}\int d^3x\, d^2\th ~ W^2~~,~~ W^2~:=~ \frac{1}{2}\,W^{\a}\,W_{\a}~~,
\ee
and the corresponding spacetime action is
\begin{equation}\label{SMc}
S ~=~ \frac{1}{g^2}\,\int d^3x~\Big\{ i\,\lambda^{\a}\,\pa_{\a}{}^\b\,\lambda_{\b} ~-~\frac12\,f^{\a\b}f_{\a\b}\Big\}~~,
\end{equation}
where the component fields $\lambda_\a$ and $f_{\a\b}$ are defined as $\lambda_\a = W_{\a}|$ and
$f_{\a\b} = \D_\a W_\b |$ respectively and $g$ is a dimensionless constant.
Due to constraint \eqref{MBI}, it is straightforward to see that the component field $f_{\a\b}$ is symmetric
in the two spinorial indices ($f_{\a\b}=f_{\b\a}$) and is the spinor form of the usual 3D Faraday tensor:
\be
F_{ab}\sim\epsilon_{abc}(\gamma^c)^{\a\b}\,f_{\a\b}~~.
\ee
% with the property $F_{ab}F^{ab}~=~f_{\a\b}f^{\a\b}$.
% In terms of $F_{\ab}$ the component Lagrangian takes the more familiar form:
% \begin{equation}
%     \begin{split}
%         \mathcal{L}^{3D}_{Maxwell} % ~=~ % \frac12\,T
%       %  ~=&~ \frac{i}{2}\,\lambda^{\a}\,\pa_{\a}{}^{\b}\,\lambda_{\b} ~-~\frac14\,f^{\a\b}f_{\a\b}\\
%         ~=&~ \frac{i}{2}\,\lambda^{\a}\,\pa_{\a}{}^{\b}\,\lambda_{\b} ~-~\frac14\, F_{\un a\un b}\,F^{\un a\un b}
%     \end{split}
%      \label{equ:fab}
% \end{equation}

\subsection{Maxwell-Goldstone Multiplet}

Our aim is to interpret the above supermultiplet as the Goldstone multiplet that accommodates the Goldstino
associated with the spontaneous breaking of the second supersymmetry in 3D.  Following \cite{Bagger:1996wp},
we search for the most general transformation $\d^*$ of $W_\a$ which is consistent with constraint
\eqref{MBI}. In order for this transformation to be understood as a second supersymmetry transformation, it
must involve the second supersymmetry partners of $W_\a$.  Due to the spontaneous breaking of this
supersymmetry, one of the partners will acquire a non-trivial VEV which will generate the characteristic shift
in the transformation of $W_\a$.  After redefining the second supersymmetry parameter and the remaining
partner superfield, we find the most general transformation is
\bea{l}\n\label{equ:eta_trans}
    \d^*_{\eta}\,W_{\a} ~=~ \eta_{\a} ~-~ \frac{1}{2\kappa} \,({\rm D}^{\b}\,{\rm D}_{\a}\,X)\,\eta_{\b}~~,\sn\label{MetaW}\\
    \d^*_{\eta}\,X ~=~ \frac{2}{\kappa}\,\eta^{\a}\,W_{\a}~~.\sn\label{MetaX}
\eea
The dimensionful parameter $\kappa$ $([\kappa]=3/2)$ corresponds to the non-trivial VEV and the overall
coefficient in \eqref{MetaX} is determined by the compatibility of \eqref{equ:eta_trans} with supersymmetry
algebra\footnote{We follow the conventions of \emph{Superspace} \cite{Gates:1983nr}. For details see appendix
\ref{3DSS}.}.

Furthermore, in order to prohibit the remaining partner superfield $X$ to carry independent degrees of freedom
and have the 3D Maxwell multiplet be the Goldstone multiplet corresponding to the partial breaking of
supersymmetry, we impose the following nonlinear constraint:
\begin{equation}\label{MXcon}
    \kappa X ~=~ W^{\a}W_{\a}~+~\frac{1}{2}\,({\rm D}^2\,X)\,X ~~\Rightarrow~~
    X~=~\frac{W^\a\,W_\a}{\kappa\,\Big(1-\frac{1}{2\kappa}\,\D^2X\Big)}~~.
\end{equation}
This constraint is determined uniquely by its compatibility with transformations \eqref{equ:eta_trans}.
Using $W_\a W_\b W_\g =0$, constraint \eqref{MXcon} can be solved in a similar fashion as in
\cite{Bagger:1996wp} and express $X$ in terms of $W^2$ and $D^2W^2$:
\be\label{MXsolution}
X~=~\frac{2}{\kappa}\,W^2\,\Bigg[\,1+\frac{T}{1-T+\sqrt{1-2T}}\,\Bigg]~~,
\ee
where $T=\frac{2}{\kappa^2}\D^2W^2$.  The above solution can be used to write the superspace action for the
3D Maxwell-Goldstone multiplet:
\be\label{SMG}
S~=~\t\,\int d^3x\,d^2\th~X ~=~ \frac{2\t}{\kappa}\,\int d^3x\,d^2\th~W^2\,\Bigg[\,1+\frac{T}{1-T+\sqrt{1-2T}}\,\Bigg]~~.
\ee
This is manifestly invariant under the first supersymmetry and also respects the second supersymmetry
because due to \eqref{MetaX} and \eqref{MBI} the spacetime Lagrangian transforms as a total derivative
\begin{equation}
        \d^*_{\eta}\,{\rm D}^2\,X ~=~ \frac{2i}{\kappa}\,\pa_{\a}{}^\b( \eta^{\a}\,W_{\b} )~~.
\end{equation}
It is clear that \eqref{SMG} is a supersymmetric Born-Infeld type of action and we will show its bosonic part
is the specific BI effective action associated with D2-brane. The superspace effective action \eqref{SMG} and
constraint \eqref{MXcon} match the results found in \cite{Ivanov:2001gd}.

\subsection[Nilpotent N=2 superfield]{Nilpotent superfield description based on $\N=2$ vector multiplet}
Using the constrained superfield approach \cite{Rocek:1978nb,Rocek:1997hi}, the above results can be
understood from the view point of partial supersymmetry breaking of a manifestly $\N=2$ theory. The 3D, $\N=2$
vector multiplet is described by a scalar superfield $\W(x,\th,\tth)$ which satisfies the following
irreducibility conditions:
\be\label{N2Mc}
\D^2\,\W ~=~ \tD^2\,\W~~,~~\D^\a\,\tD_\a\,\W ~=~ 0~~,
\ee
where the tilded Grassmann coordinates ($\tth_\a$) and covariant derivatives ($\tD_\a$) correspond to the
second supersymmetry. By expanding the $\N=2$ superfield $\W$ in terms of $\N=1$ component superfields
\be
\W(x,\th,\tth)~=~ \Phi(x,\th)~+~\tth^\a\, W_\a(x,\th)~-~ \tth^2\,F(x,\th)~~,
\ee
we can solve \eqref{N2Mc} to find that
\be\label{N2Mcc}
F~=~\D^2\,\Phi~~,~~\D^2\,W_\a~=~i\pa_\a{}^\b\,W_\b~~,~~\D^\a\,W_\a~=~0~~.
\ee
Moreover, superfields $\Phi$, $W_\a$, and $F$ transform under the second supersymmetry as follows:
\bea{l}\n\label{N2Msusy}
\d^*_\e\,\Phi ~=~ -\, \e^\a\,W_\a~~,\sn\\
\d^*_\e\,W_\a ~=~ \e_\a\,F~-~i\,\e^\b\,\pa_{\b\a}\,\Phi ~=~ \D^\b\D_\a\,\Phi\,\e_\b~~,\sn\\
\d^*_\e\,F~=~ -i\,\e^\a\,\pa_\a{}^\b \,W_\b~~,\sn
\eea
which are consistent with \eqref{N2Mcc}. In order to break this second supersymmetry we
(\emph{i}) expand the $\N=2$ superfield $\W$
around a background superfield which breaks the second supersymmetry and (\emph{ii}) we impose the usual
nilpotence condition in order to remove `radial' superfields:
\be
\W~=~\langle\W\rangle~+~\eW~~,~~\eW^2~=~0~~.
\ee
The condensate $\langle\W\rangle$ is Lorentz and $\N=1$ invariant with a non-trivial $\tth$ dependence.
\be
\langle\W\rangle ~=~ \kappa~\tth^2~~\Rightarrow~~\eW~=~-\frac{1}{2}\,X~+~\tth^\a\,W_\a-\tth^2\,\left(-\frac{1}{2}\,\D^2\,X+\kappa\right) ~~,
\ee
where $X=-2\Phi$. The second supersymmetry transformations take the form
\bea{l}\n
\d^*_\e\,X~=~2\, \e^\a\,W_\a~=~\frac{2}{\kappa}\, \pmb{\e}^\a\,W_\a~~,\sn\\
\d^*_\e\,W_\a~=~\e_\a\,\kappa ~-~\frac12\,(\D^\b\D_\a\,X)\,\e_\b~=~\pmb{\e}_\a ~-~\frac{1}{2\kappa}\,(\D^\b\D_\a\,X)\,\pmb{\e}_\b~~,\sn
\eea
where $\pmb{\e}_\a$ is the $\kappa$ scaled supersymmetry parameter (~$\pmb{\e}_\a=\e_\a\,\kappa$~). These
transformations match exactly \eqref{equ:eta_trans}.  The nilpotence condition $\eW^2=0$ imposes the following
nonlinear constraints:
\be
X^2~=~0~,~X\,W_\a~=~0~~,~~\kappa\,X~=~W^\a\,W_\a~+~\frac12\,X\D^2X~~.
\ee
which generates \eqref{MXcon}.

\subsection{Supersymmetric 3D Born-Infeld action in components}
Starting from the superspace action \eqref{SMG}, we extract the spacetime component action.  It can be written
in the following way
\be
S~=~ \t\,\kappa\,\int d^3x~\Bigg\{T\big\rvert_{\th=0}~+~\frac{2}{\kappa^2}\int d^2\th~\Psi(T)\,W^2\Bigg\}~~,
\label{equ:sBI-action}
\ee
which makes it easy to see that it is a member of the Cecotti-Ferrara class of actions \cite{Cecotti:1986gb}
after a dimensional reduction to 3D\footnote{For details look in appendix \ref{CF}.}. In this case,
the function $\Psi(x)$ is fixed to be:
\be\label{Psi}
\Psi(x)~=~\frac{x}{1-x+\sqrt{1-2x}}~~.
\ee
By performing the $\th$ integral we find the bosonic part of the spacetime action to be
\begin{equation}
    \begin{split}
        S_{\rm B}
 ~=&~ \tau\,\kappa \,\int\,d^3x\, \Big(\,1 ~-~ \sqrt{1-2s}\,\Big) ~~,
    \end{split}
    \label{equ:BI}
\end{equation}
where
\begin{equation}
    s ~=~ -\frac{1}{\kappa^2}\,f^{\a\b}\,f_{\a\b}~~,
\end{equation}
and corresponds to the 3D BI effective action.
The fermionic part of the spacetime action is:
\bea{l}\n\label{FLM}
S_{\rm F}~=~ \frac{\tau}{\kappa^3}\,\int d^3x~\Bigg\{\Psi'(T|)\,\Bigg[~
4i\,(f^{\a\d}\,\lambda_{\d})~\pa_{\a\b}\,(\,f^{\b\g}\,\lambda_{\g})\\[-1mm]
\hspace{46mm}+\,\bigg(2~(\Box\,\lambda^{\g})~\lambda_{\g} ~+~ (\pa^{\a\b}\lambda^{\g})\,(\pa_{\a\b}\lambda_{\g})\bigg)~
\lambda^{\s}\,\lambda_{\s}~\Bigg]\\
\hspace{31mm}~-\frac{2}{\kappa^2}\,\Psi''(T|)\,
\Big[\pa^{\a}{}_{\b}(f^{\b\g}\lambda_{\g})\Big]\,\Big[\pa_{\a}{}_{\d}(f^{\d\e}\lambda_{\e})\Big]~\lambda^{\s}\,\lambda_{\s}
\Bigg\}~~,
\eea
where
\begin{align}
    & T| ~=~ \frac{2}{\kappa^2} \Bigg[\,-\frac{1}{2}\,f^{\a\b}f_{\a\b} ~-~ i\,\lambda^{\a}\,(\pa_{\a\b}\lambda^{\b})\,\Bigg]~~,\\
    &\Psi'(T|) ~=~ \frac{1}{\sqrt{1-2T|}}\,\frac{1}{1-T| + \sqrt{1-2T|}}~~, \\
     &\Psi''(T|) ~=~ \frac{2-3T|+2\sqrt{1-2T|}}{(1-2T|)^{\frac{3}{2}}(1-T|+\sqrt{1-2T|})^2}~~.
\end{align}
The structure of the fermionic part of the Lagrangian is much more complicated and has the form of linear
combination of polynomial terms weighted by non-polynomial factors that correspond to the first and second
derivatives of the 3D Ceccoti-Ferrara function $\Psi(x)$.

\section{Gauge invariant Chern-Simons Mass}\label{sec:deformation}
A special property of the 3D vector multiplet which makes it very different from its 4D analogue is the
existence of a gauge invariant mass term
\be\label{Smass}
S_m~=~\frac{m}{2}\,\int d^3x\,d^2\th~\Gamma^\a\,W_\a ~~.
\ee
The gauge invariance of this term is based on the Bianchi identity \eqref{MBI} generated by varying
\eqref{Smass} via \eqref{equ:gauge-Gamma}. This term is manifestly invariant under the first, linear realized,
supersymmetry and we want to study its transformation under the second supersymmetry. That requires to have
knowledge of the transformation properties of the prepotential $\Gamma_\a$ under the second supersymmetry and
so for its supersymmetric partner. With that in mind
We introduce a new superfield $\Delta_\a$ defined as follows
\be\label{Delta}
\D^\a\,\Delta_\a~=~X~~.
\ee
This definition does not determine $\Delta_\a$ uniquely, as it enjoys the gauge transformation
\be\label{gaugeDelta}
\d\Delta_\a~=~\D^\b\,\D_\a\,K_\b~~,~~\d K_\a~=~\D_\a\,K~~.
\ee
Using \eqref{W} and \eqref{Delta}, we find that transformations \eqref{equ:eta_trans} induce the following
second supersymmetry transformations for superfields $\Gamma_\a$ and $\Delta_\a$
\bea{l}\n\label{dsusyGD}
\d^*\Gamma_\a~=~\frac{1}{\kappa}\,\eta^\b\,\D_\b\,\Delta_\a~+~\Phi^{(sp)}_\a~~,\sn\label{dsusyG}\\
\d^*\Delta_\a~=~-\frac{1}{\kappa}\,\eta^\b\,\D_\b\,\Gamma_\a~~,\sn
\eea
where $\Phi^{(sp)}_\a=-2\eta_\a\,\th^2$ is the special solution of the inhomogeneous equation
$\D^\b\,\D_\a\,\Phi^{(sp)}_\b$=$2\eta_\a$. The full solution of the homogeneous equation
$\D^\b\,\D_\a\,\Phi_\b=0$ corresponds to a gauge transformation of $\Gamma_\a$ and thus can be dropped since
\eqref{dsusyG} is valid modulo gauge transformation terms. Under the above transformations \eqref{dsusyGD},
the mass term \eqref{Smass} is not invariant
\be
\d^*\,S_m~=~\frac{m}{g^2}\,\int d^3x\,d^2\th~\Gamma^\a\,\eta_\a~=~\frac{2m}{g^2}\int d^3x~\l^\a\,\eta_\a~~,
\ee
where $\D^2\Gamma_\a\big\rvert_{\th=0}=2\,W_\a\big\rvert_{\th=0}=2\,\l_\a$.  However, it is easy to check that
the integrand $2\,\l^\a\,\eta_\a$ corresponds to the second supersymmetry transformation \eqref{MetaX} of
$X\big\rvert_{\th=0}$.  Therefore, there exists a one-parameter family of deformations of the
Maxwell-Goldstone action
\be\label{Sxi}
S_{\xi}~=~\frac{\xi\,m}{2}\,\int d^3x\,d^2\th~\Big\{\Gamma^\a\,W_\a ~-~ 2\kappa\,\th^\a\,\Delta_\a\Big\}~~,
\ee
which respects the second supersymmetry transformations \eqref{dsusyGD} and gauge transformations
\eqref{equ:gauge-Gamma} and \eqref{gaugeDelta}.
Nevertheless, the second term of $S_\xi$
\be\label{SX}
\int d^3x d^2\th~ \th^\a\Delta_\a ~=~ \int d^3x\, X\big\rvert_{\th=0}
\ee
explicitly breaks the first supersymmetry. Using \eqref{MXsolution}, it is straightforward to find that
\be
X\big\rvert_{\th=0}~=~\frac{1}{\kappa}\,\l^\a\l_\a\,\Big(1+\Psi(T\big\rvert)\Big)~~,
\ee
which does not include a bosonic part, thus it can not be invariant under the first
supersymmetry transformation.

\section{3D, ${\cal N}=1$ Tensor-Goldstone Multiplet}
\label{sec:TG}

In four dimensions, it has been shown that chiral multiplets, vector multiplets, or tensor multiplets can play
that role \cite{Bagger:1994vj,Bagger:1996wp,Bagger:1997pi,Rocek:1997hi} of Goldstone multiplets that can be
used to accommodate the goldstion field generated by the partial supersymmetry breaking. In three dimension of
course there is no notion of $\N=1$ chirality, hence we explore the use of the 3D reduction of the tensor
multiplet.

\subsection{Review of 3D, ${\cal N}=1$ Tensor multiplet}

% In three dimensions, the tensor multiplet is described by a spinorial superfield $U_\a$ which satisfies the
Consider a spinorial superfield $U_{\a}$ which satisfies the following constraint:
% following constraint:
\be\label{TMcon}
\D^\a\,\D_\b\,U_\a~=~0~~\Rightarrow~~\D^2\,U_\a~=~-i\,\pa_\a{}^\b\,U_\b~~.
\ee
This can be solved by expressing $U_\a$ in terms of an unconstrained scalar superfield $G$:
\be
U_\a~=~\D_\a\,G~~.
\ee
Superfield $G$ can be understood as the field strength of a spinorial gauge superfield $\Psi_{\a}$
\be
G=\D^{\a}\Psi_{\a}
\ee
with the following gauge transformation
\be
\delta\Psi_{\a}=\D^{\b}\D_{\a}K_{\b}~,~\delta K_{\a}=\D_{\a}K~.
\ee
This supermultiplet corresponds to the 3D reduction of the 4D Tensor multiplet and we will refer to it as the
3D tensor multiplet.
The superspace action describing the free dynamics of tensor multiplet takes the form
\be\label{Stm}
S~=~-\frac{1}{g^2}\,\int d^3x\,d^2\th~U^2~~,
\ee
which generates the spacetime action
\begin{equation}
    \begin{split}
        S ~=&~ \frac{1}{g^2}\,\int d^3x\,\Bigg\{\,-\,\frac{1}{2}\,(\pa^{\a\b}\,\phi)(\pa_{\a\b}\,\phi)~+~  H^2 ~-~ i\, \chi^{\a}\,(\pa_{\a\b}\,\chi^{\b})  \,\Bigg\}\\
        ~=&~ \frac{1}{g^2}\,\int d^3x\,\Bigg\{\, \frac{1}{2}\,\hat{f}^{\a\b}\hat{f}_{\a\b}~-~ i\, \chi^{\a}\,(\pa_{\a\b}\,\chi^{\b})  \,\Bigg\}~~,
    \end{split}
\end{equation}
where we define the component fields by the following projection
\begin{equation}
\begin{split}
   & U_{\a}| ~=~  \chi_{\a}~~~,~~~\D_{\a}\,U_{\b}| ~:=~ \hat{f}_{\a\b} ~=~ i\,\pa_{\a\b}\,\phi -\,C_{\a\b}\,H  ~~~,~~~
   \D^2\,U_{\a}| ~=~  i\,\pa_{\a\b}\,\chi^{\b}~~~.
\end{split}
\label{equ:U-proj}
\end{equation}
Notice that the component $\D_{\a}\,U_{\b}|$, in contrast to $\D_{\a}W_{\b}|$ of \eqref{SMc},
is not symmetric in the two spinorial indices.
Therefore it can be decomposed into a symmetric and anti-symmetric part labeled by component
fields $\phi$ and $H$ respectively which are defined as follows:
\begin{equation}
    G| ~=~ \phi ~~,~~ \D^2\,G| ~=~ H~~.
\end{equation}

Furthermore, in 3D, tensor and vector multiplets are related by a duality transformation. This can be easily
seen by the following parent action
\be\label{MTD}
S~=~\frac{1}{g^2}\,\int d^3x\,d^2\th ~\Big\{W^2 + \L^\a\,\big(\,W_\a-\frac12\,\D^\b\,\D_\a\,\Gamma_\b\,\big)\Big\}~~,
\ee
where superfields $W_\a,~\Gamma_\a$ and $\L_\a$ are unconstrained. The Lagrange multiplier $\L_\a$, once
integrated out, it identifies $W_\a$ with the vector multiplet superfield strength \eqref{W} and enforces
constraints \eqref{MBI}.  As a result we get the free vector multiplet action \eqref{Svm}. On the other hand,
by integrating out superfield $\Gamma_\a$ first, we get that the superfield $\L_\a$ satisfies the following
equation of motion
\be\label{MTDc}
\D^\a\,\D_\b\,\L_\a ~=~ 0~~,
\ee
hence $\L_\a$ becomes a tensor multiplet superfield. Finally, by integrating out the superfield $W_\a$, we
find its equation of motion to be $W_\a=-\L_\a$ which when substituted back gives the superspace action
\eqref{Stm}.

\subsection{Tensor-Goldstone Multiplet}

Interpreting the 3D tensor multiplet as the Goldstone multiplet corresponding to the breaking of the second
supersymmetry requires to find a transformation of $U_\a$ compatible with the constraint \eqref{TMcon} which
includes a constant shift term and has the interpretation of supersymmetry ---must involve partners and be
consistent with the susy algebra. The most general transformation of this type is:
\bea{l}\n\label{etaT}
    \d^*_{\eta}\,U_{\a} ~=~ \eta_{\a} ~-~ \frac{1}{2\tilde{\kappa}}\,(\D_{\a}\D^{\b}\Tilde{X})\,\eta_{\b}~~,\sn \\
    \d^*_{\eta}\,\Tilde{X} ~=~  \frac{2}{\tilde{\kappa}}\,\eta^{\a}\,U_{\a}~~.\sn\label{TetaX}
\eea
Notice that \eqref{etaT} and \eqref{equ:eta_trans} differ in the order in which the spinorial covariant
derivatives act on the partner superfield. In order to remove the independent degrees of freedom in
$\tilde{X}$, we impose a non-linear constraint which expresses $\tilde{X}$ as a function of $U_\a$ and its
derivatives. The compatibility of this constraint with \eqref{etaT} determines it to be the following:
\be\label{TXcon}
\tilde{\kappa}\tilde{X} ~=~ U^\a\,U_\a-\frac12\,(\D^2\,\tilde{X})\,\tilde{X}~~\Rightarrow~~
\tilde{X}~=~\frac{U^\a\,U_\a}{\tilde{\kappa}\Big(1+\frac{1}{2\tilde{\kappa}}\D^2\tilde{X}\Big)}~~.
\ee
Similar to \eqref{MXcon}, using $U_\a\,U_\b\,U_\g=0$, this constraint can be solved in order to express
$\tilde{X}$ in terms of $U^2$ and its derivative $\tilde{T}=\frac{2}{\tilde{\kappa}^2}\,\D^2U^2$:
\begin{equation}
    \Tilde{X} ~=~ \frac{2}{\tilde{\kappa}}\, U^2 \left[ 1- \frac{\Tilde{T}}{1+\Tilde{T}+\sqrt{1+2\Tilde{T}}} \right] ~~~.
    \label{equ:X-ansatz-GT}
\end{equation}
The superspace action for the 3D Tensor-Goldstone multiplet is
\be\label{STG}
S~=~ -\,\tilde{\t}\int d^3x\,d^2\th~\tilde{X}~=~-\frac{2\tilde{\t}}{\tilde{\kappa}}\,\int d^3x\,d^2\th~U^2\,\left[1-\frac{\tT}{1+\tT+\sqrt{1+2\tT}}\right]~~.
\ee
It is manifestly invariant under the first supersymmetry and it is straightforward to check its invariance
under the second supersymmetry.  Due to \eqref{TetaX} and \eqref{TMcon} the spacetime Lagrangian transforms
under the second supersymmetry as a total derivative
\begin{equation}
        \d^*_{\eta}\,\D^2\tX ~=~ -\frac{2i}{\tilde{\kappa}}\,\pa_{\a}{}^\b( \eta^{\a}\,U_{\b} )~~.
\end{equation}
The action \eqref{STG}, as well as the constraint \eqref{TXcon}, match the results of \cite{Ivanov:1999fwa}.

\subsection[Nillpotent N=2 superfield]{Nillpotent superfield description based on $\N=2$ chiral multiplet}
We now show that these results emerge from the partial supersymmetry breaking procedure of the 3D, $\N=2$
chiral multiplet. Consider an ${\cal N}=2$ scalar superfield $\U(x,\th,\tth)$ which satisfies the
following conditions:
\be\label{Uconst}
\D^2\U~=~-\,\tD^2\U~,~\D^\a\,\D_\b\,\tD_\a\,\U=0~~.
\ee
We solve these constraints by expanding the $\N=2$ superfield $\U$ in its $\N=1$ superfield components
\be
\U(x,\th,\tth)~=~\Phi(x,\th)~+~\tth^\a\, U_\a(x,\th)~-~\tth^2\,F(x,\th)~~,
\ee
which satisfy the following relations:
\be
F~=~-\,\D^2\Phi~~,~~\D^2\, U_\a~=~-i\,\pa_\a{}^\b\,U_\b~~,~~\D^\a\,\D_\b\,U_\a~=~0~~.
\ee
Their transformations under the second supersymmetry are
\bea{l}\n\label{N2Tsusy}
\d^*_\e\,\Phi~=~-\, \e^\a\,U_\a~~,\sn\\
\d^*_\e\,U_\a~=~\e_\a\,F-i\,\e^\b\,\pa_{\b\a}\,\Phi\,=\D_\a\D^\b\,\Phi\,\e_\b~~,\sn\\
\d^*_\e\,F~=~-i\,\e^\a\,\pa_\a{}^\b \,U_\b~~.\sn
\eea
Constraints \eqref{Uconst} can also be solved by considering an isodoublet $(\U,\V)$ of $\N=2$ scalar superfields
which satisfy the conditions:
\begin{equation}
    \tD_{\a}\,\U = -\D_{\a}\,\V~,~\tD_{\a}\,\V = \D_{\a}\,\U
\end{equation}
These conditions can be combined to give
\begin{equation}
    \Big(\tD_{\a}+i\D_{\a}\Big)\Big(\U-i\V\Big)=0
\end{equation}
which is the covariantly chiral condition for the superfield $\Phi=\U-i\V$. Therefore superfield $\U$
corresponds to the real part of the $3D,~\N=2$ chiral superfield $\Phi$ used in
\cite{Ivanov:1999fwa,Ivanov:2001gd}.

Applying on this supermultiplet the constrained superfield approach in order to break the second manifest
supersymmetry, we expand $\U$ around a non-trivial vacuum that preserves only the first supersymmetry and at
the same time we eliminate the remaining partner superfield by imposing the nilpotency condition:
\be
\U~=~ \langle\U\rangle~+~\eU~~,~~\eU^2~=~0~~.
\ee
The condensate $\langle\U\rangle$ is Lorentz and $\N=1$ invariant with a non-trivial $\tth$ dependence.
\be
\langle\U\rangle~=~ \tilde{\kappa}~\tth^2~~\Rightarrow~~
\eU~=~-\frac{1}{2}\,\tX~+~\tth^\a\,U_\a~-~\tth^2\,\left(\frac{1}{2}\,\D^2\,\tX+\tilde{\kappa}\right)
\ee
where $\tX=-2\Phi$. The second supersymmetry transformations take the form
\bea{l}\n
\d^*_\e\,\tX~=~2\, \e^\a\,U_\a~=~\frac{2}{\tilde{\kappa}}\, \pmb{\e}^\a\,U_\a~~,\sn\\
\d^*_\e\,U_\a~=~\e_\a\,\tilde{\kappa} ~-~\frac12\,(\D_\a\,\D^\b\,\tX)\,\e_\b~=~\pmb{\e}_\a ~-~\frac{1}{2\tilde{\kappa}}\,(\D_\a\,\D^\b\,\tX)\,\pmb{\e}_\b~~,\sn
\eea
where $\pmb{\e}_\a$ is the $\tilde{\kappa}$-rescaled supersymmetry parameter and are in complete agreement
with transformations \eqref{etaT}. Moreover,
the nilpotence condition $\eU^2=0$ generates the nonlinear constraint \eqref{TXcon}:
\be
\tX^2~=~0~,~\tX\,U_\a~=~0~~,~~\tilde{\kappa}\,\tX~=~U^\a\,U_\a~-~\frac12\,(\D^2\tX)\,\tX~~.
\ee

\subsection{Spacetime action for Tensor-Goldstone multiplet}
The superspace action \eqref{STG} can be written in the three-dimensional Ceccoti-Ferrara form
\be
S~=~-\,\tilde{\t}\,\tilde{\kappa}\,\int d^3x~\Bigg\{\Tilde{T}\big\rvert_{\th=0}~+~\frac{2}{\tilde{\kappa}^2}\int d^2\th~{\Psi}(-\Tilde{T})\,U^2\Bigg\}~~.
\ee
where $\Psi(x)$ is the same function as in the Maxwell-Goldstone case \eqref{Psi}. By performing the
grassman integral we extract the bosonic part of the spacetime action:
\be\label{BLT}
S_{\rm B}~=~\tilde{\t}\,\tilde{\kappa}\,\int d^3x\,\Big(\, 1 - \sqrt{1+2\,\Tilde{T}|}\,\Big)~~.
\ee
The fermionic part of the action is
\begin{equation}\label{FLT}
    \begin{split}
        S_{\rm F} ~=&~\frac{\tilde{\t}}{\tilde{\kappa}^3}\,\int d^3x\,\Bigg\{\, \Psi'(-\Tilde{T}|)\,\Bigg[\, 4i\,\pa_{\a\b}(\hat{f}^{\b\g}\,\chi_{\g})\,\hat{f}^{\a\d}\,\chi_{\d} \\
         &\qquad\qquad\qquad\qquad\qquad~+~ \bigg(2~(\Box\,\chi^{\g})~\chi_{\g} ~+~ (\pa^{\a\b}\chi^{\g})\,(\pa_{\a\b}\chi_{\g})\bigg)\,\chi^{\s}\chi_{\s} \,\Bigg]\\
        &\hspace{23mm}+~ \frac{2}{\tilde{\kappa}^2}\,\Psi''(-\Tilde{T}|)\,\Big[\pa^{\a}{}_{\b}(f^{\b\g}\lambda_{\g})\Big]\,\Big[\pa_{\a}{}_{\d}(f^{\d\e}\lambda_{\e})\Big]\,\chi^{\s}\chi_{\s}\,\Bigg\}~~,\\
    \end{split}
\end{equation}
where
\begin{equation}
    \Tilde{T}| ~=~ \frac{2}{\tilde{\kappa}^2} \Bigg[\,-\frac{1}{2}\,\hat{f}^{\a\b}\hat{f}_{\a\b} ~+~ i\,\chi^{\a}\,(\pa_{\a\b}\chi^{\b})\,\Bigg]~~.
\end{equation}

\section{Tensor-Goldstone and Maxwell-Goldstone Duality}\label{sec:duality}

In 4D, the self-duality of the Born-Infeld action was extended to the supersymmetric BI action based on the
self-duality of the Maxwell-Goldstone multiplet. However, ss demonstrated in \eqref{MTD}, in 3D the vector
multiplet is no longer self-dual but it maps to the tensor multiplet. We will show that this duality survives
between the Maxwell-Goldstone and Tensor-Goldstone multiplets.

A very transparent method for studying the duality properties of these multiplets is\footnote{See
\cite{Tseytlin:1996it,Rocek:1997hi}.} to consider unconstrained superfields and impose all nonlinearities and
constraints via Lagrange multipliers. Therefore, the Maxwell-Goldstone action \eqref{SMG}
can be written in the form
\begin{equation}\label{SMGL}
 S ~=~ \int\,d^3x\,d^2\theta~\Bigg\{\Lambda\,\left[\,W^{\a}W_{\a} ~+~ \frac{1}{2}\,X\,\D^2\,X ~-~ \kappa\, X\,\right]
 ~+~ \t\, X\Bigg\}~~,
\end{equation}
where $\L$ and $X$ are unconstrained scalar superfields. When the Lagrange multiplier $\L$ is integrated out,
it imposes the susy breaking constraint \eqref{MXcon} and the above action becomes identical to \eqref{SMG}.
Action \eqref{SMGL} can also be motivated by the free $\N=2$ action (the sum of kinetic energy terms for
$W_\a$ and $X$) plus a constraint term with a Lagrange multiplier. Furthermore, using \eqref{MTD} and
\eqref{MTDc}, we can relax the $\D^\a\,W_\a=0$ constraint of the vector multiplet $W_\a$ by adding a duality
term with a Lagrange multiplier which must be a tensor supermultiplet spinorial superfield $U_\a$
($\D^\a\,\D_\b\,U_\a=0$)
\be
S ~=~ \int\,d^3x\,d^2\theta~\Bigg\{\Lambda\,\left[\,W^{\a}W_{\a} ~+~ \frac{1}{2}\,X\,\D^2\,X ~-~ \kappa\, X\,\right]
 ~+~ \t\, X ~+~ g\,U^\a\,W_\a\Bigg\}~~.
\ee
Integrating out $U_\a$ restores the vector constraint $\D^\a\,W_\a=0$. However, because $W_\a$ is now
unconstrained and appears algebraically, we can choose to integrate it out first. The result is the following
action
\begin{equation}\label{Sdual}
    S ~=~ \int\,d^3x\,d^2\theta~\Bigg\{\Tilde{\Lambda}\,\Bigg[\,U^{\a}U_{\a} ~-~ \frac{1}{2}\,\Tilde{X}\,\D^2\,\Tilde{X} ~-~ \Tilde{\kappa}\, \Tilde{X}\Bigg] ~-~ \Tilde{\tau}\, \Tilde{X}
\,\Bigg\}~~,
\end{equation}
where
\begin{equation}\label{dual-parameter}
    \Tilde{\Lambda} ~=~ -\frac{g^2}{4\,\Lambda} ~~,~~
    X~=~ \frac{g\,\Tilde{X}}{2\,\Lambda}~~,~~
    \Tilde{\kappa} ~=~ \frac{2\,\tau}{g}~~,~~
    \Tilde{\tau} ~=~ \frac{g\,\kappa}{2}~~.
\end{equation}
Action \eqref{Sdual} corresponds to the Tensor-Goldstone multiplet action \eqref{STG}, since the Lagrange
multiplier $\tilde{\L}$ imposes constraint \eqref{TXcon}. Moreover, the relation between the parameters
$\kappa, \t$ that appear in Maxwell-Goldstone action
and the corresponding Tensor-Goldstone parameters $(\tilde{\kappa}, \tilde{\t})$ is
such that the dimensionless parameter $\lambda=\frac{\kappa}{\tau}$ undergoes a standard inversion
\be\label{l}
\lambda\to\tilde{\lambda}=\frac{4}{g^2\,\lambda}~.
%\t\,\kappa~=~\tilde{\t}\,\tilde{\kappa}~~.
\ee
This is reminiscent of the inversion included in the SL(2,R) duality transformation of the generalized BI action
obtained by turning on the various background fields.  Furthermore, notice that this duality can be extended
to the corresponding $\N=2$ multiplets.  Specifically, if one chooses the Lagrange multiplier $\Lambda$ to be
the constant $\Lambda=\frac{\tau}{\kappa}$, then the linear $X$ terms in \eqref{SMGL} drop and we recover the
free $\N=2$ multiplet written in terms of $\N=1$ superfields.  Under the map \eqref{dual-parameter}
$\tilde{\Lambda}=-\frac{\tilde{\tau}}{\tilde{\kappa}}$ which also lead to the cancellation of the linear terms
in  \eqref{Sdual} and thus describing the the free $\N=2$ tensor multiplet.

\section{Summary}
The existence of solitonic, BPS, solutions (D$p$-branes) of type II string theory motivates the study of
supersymmetric extensions of Born-Infeld type actions. These are viewed as low-energy effective descriptions
---hence they are not subject to any renormalizability requirements--- which correctly capture the spontaneous
breaking of half of the supersymmetries. Therefore, these effective supersymmetric actions can be written in
terms of linear representations of $\N=1$ supersymmetry and are also invariant under a second, nonlinear
supersymmetry transformation.
% It has long been understood that broken symmetries lead to nonlinear
% representations of the broken symmetry, which accommodate the Goldstone modes.
%
In 4D, such effective supersymmetric BI actions have been constructed
\cite{Cecotti:1986gb,Bagger:1994vj,Bagger:1996wp,Bagger:1997pi,Rocek:1997hi} and studied extensively.  It was
shown that the Goldstone fermion, corresponding to the spontaneously broken second supersymmetry, could be
accommodated in an $\N=1$ chiral, vector, or tensor multiplets. For each one of such descriptions, the $\N=1$
manifestly supersymmetric action was constructed and showed that the bosonic part of these actions matched the
expected BI action. Moreover, it was shown that the Maxwell-Goldstone multiplet is self-dual and the Tensor
and Chiral Goldstone multiplets map to each other under duality transformations. At the component level, these
duality properties reproduced the  self-duality of the BI action. Finally, it was later shown
\cite{Antoniadis:2008uk} that the requirement of invariance under a first linear supersymmetry and a second
nonlinear supersymmetry does not uniquely determine the action. The addition of a FI term preserves the
nonlinearly realized supersymmetry,  but it spontaneously breaks the linear supersymmetry. However, the
deformed theory still describes a partial and not full supersymmetry breaking, with the surviving
supersymmetry to correspond to a different $\N=1$ slice of the $\N=2$ theory.

In this work, motivated by the special role of D2-brane in type IIA string theory, we aim towards the
construction of 3D supersymmetric Born-Infeld actions. Effective actions of this type have been obtained by
performing dimensional reductions from 4D to 3D which break half of the supersymmetries
\cite{Ivanov:1999fwa,Ivanov:2001gd}. In contrast, we consider manifestly supersymmetric $\N=2, D=3$ multiplets
with no central charges and we decompose them to their $\N=1$ constituents.  Next, we expand around a
nontrivial vacuum that breaks the second supersymmetry and at the same time we enforce the nilpotence
conditions \`{a} la \cite{Rocek:1978nb} in order to (\emph{a}) eliminate additional degrees of freedom present
in the $\N=2$ multiplet and (\emph{b}) give the Goldstone property to the surviving $\N=1$ multiplet.
Specifically, we find two such $\N=2$ multiplets, the vector and the chiral which give rise to a
description of the Goldstone multiplet in terms of an $\N=1$ vector or an $\N=1$ tensor multiplet
respectively. For both descriptions, we derive explicitly:\\[1mm]
(\emph{i}) the set of transformations
\eqref{equ:eta_trans} and \eqref{etaT} which satisfy the supersymmetry algebra and are consistent with the
irreducibility conditions of $W_\a$ and $U_\a$ superfields respectively\\[1mm]
(\emph{ii}) the nonlinear constraints \eqref{MXcon} and \eqref{TXcon} which define the broken supersymmetry
partner $X$ ($\tX$) of $W_{\a}$ ($U_{\a}$) as a nonlinear function of $W_{\a}$ ($U_{\a}$) in a manner
consistent with the above transformations.\\[1mm]
(\emph{iii}) the solution of the above constraints and use it to write the manifestly $\N=1$
supersymmetric extension of the 3D Born-Infeld action associated with D2-branes which is also invariant under
the nonlinear supersymmetry. \\[1mm]
These results are consistent with the results found in \cite{Ivanov:1999fwa,Ivanov:2001gd}.

Moreover, for both superspace Goldstone multiplet actions we extract the corresponding spacetime components
actions.  Their bosonic parts \eqref{equ:BI} and \eqref{BLT} match the expected 3D BI action as expected.  The
fermionic part of the Lagrangians \eqref{FLM} and \eqref{FLT} are organized into sums of polynomial terms
weighted by non-polynomial factors which correspond to first and second derivatives of the 3D Ceccoti-Ferrara
function which is identified from the superspace action.
In addition, we investigate the duality properties
of these two descriptions. We find that under duality the
Maxwell-Goldstone action \eqref{SMG} maps to the Tensor-Goldstone multiplet \eqref{STG}. This property is
inherited from the duality between 3D $\N=1$ vector and tensor multiplets and it is consistent with the
$\N=2$ viewpoint. One of the consequences of this duality is to force the inversion
\eqref{l} of the dimensionless parameter $\lambda=\frac{\kappa}{\tau}$.  Finally, we explore the possibility
of deforming the Maxwell-Goldstone superspace action by a CS-like mass term which is a characteristic term in
3D.  We find that such a term
generates a one-parameter deformation \eqref{Sxi} of the action which is consistent with the second
---nonlinearly realized--- supersymmetry, but it explicitly breaks the first, linear, supersymmetry.

As a concluding remark, we want to mention that an alternative approach to organizing nonlinearly realized
supersymmetries is the $T\bar{T}$ deformation \cite{Zamolodchikov:2004ce,Smirnov:2016lqw}.  This approach has
been studied extensively for supersymmetric and non-supersymmetric theories in two and four dimensions. The 4D
Born-Infeld action and its supersymmetric extension have been recently understood as a generalized $T\bar{T}$
deformation \cite{Conti:2018jho,Ferko:2019oyv}.  In the future, we would like to investigate if the 3D
supersymmetric BI action can have a similar interpretation. Meaning, is there an operator which depends on the
3D supercurrent and drives a flow resulting in the 3D supersymmetric BI action?

\begin{acknowledgments}
We would like to thank William D. Linch, III for his contribution during the early stages of this project. We
also would like to thank Professors E. Ivanov and S. S. Sethi for their constructive feedback on an early
version of the manuscript.  We would like to thank Jim Gates for useful conversations.  The research of Y. H.
and K.~K. is supported in part by the endowment from the Ford Foundation Professorship of Physics at Brown
University. The work of Y. H. is supported in part by the Physics Dissertation Fellowship provided by the
Department of Physics at Brown University as well.  Y.H. and K. K. gratefully acknowledge the support of the
Brown Theoretical Physics Center.  \end{acknowledgments}

\appendix

\section{Projection from 4D Cecotti-Ferrara Action}\label{CF}

In this appendix, we review the Cecotti-Ferrara Lagrangian \cite{Cecotti:1986gb} in 4D, ${\cal N}=1$ superspace, and construct the 3D Cecotti-Ferrara action via the dimensional reduction.

First, recall that the 4D, ${\cal N}=1$ Cecotti-Ferrara Lagrangian \cite{Cecotti:1986gb} reads
\begin{equation}
\begin{split}
    \mathcal{L}^{4D}_{CF} ~=&~ \hat{T} ~+~ \int\,d^2\theta\,d^2\bar{\theta}\,\Psi(T,\Bar{T})\,W^2\,\Bar{W}^2 ~~,\\
   % ~=&~ T ~+~ \Bar{T} ~+~ \fracm12\,\int\,d^2\theta\,\left[\,\Psi(T,\Bar{T})\,W^2\,T ~+~ W^2\Bar{W}(\cdots) \right]
\end{split}
\label{equ:4D-CF-pro}
\end{equation}
where
\begin{equation}
    \begin{split}
       & T ~=~ \frac{1}{2}\,\Bar{\rm D}^2\,\Bar{W}^2 ~~,~~
        \Bar{T} ~=~ \frac{1}{2}\,{\rm D}^2\,{W}^2 ~~,\\
       & \Psi(T,\Bar{T}) ~=~ \fracm{1}{1-\hat{T}+\sqrt{1-2\hat{T}-\check{T}^2}}~~,\\
       & \hat{T} ~=~ \frac{1}{2}(T+\bar{T}) ~~~,~~~ \check{T}~=~ \frac{1}{2i}(T-\bar{T})~~.
    \end{split}
\end{equation}

Before projecting the CF Lagrangian to 3D, first note that in 4D, ${\cal N}=1$ superspace, superfields are complex, the chirality and complex conjugate operation are well-defined, and there are two types of superspace covariant derivatives $\D_{\a}$ and $\bar{\D}_{\dot\a}$.
However, in 3D, ${\cal N}=1$ superspace, superfields are real, the chirality and complex conjugate operation are not defined, and there is only one type of superspace covariant derivative $\D_{\a}$.
Then, one can project the 4D Lagrangian (\ref{equ:4D-CF-pro}) to 3D
simply by (\emph{i}) doing $d^2\bar{\theta}$ integral first, $\int d^2\bar{\theta}\Psi(T,\Bar{T})\,W^2\,\Bar{W}^2 = \Psi(T,\Bar{T})\,W^2\,\bar{\D}^2\,\Bar{W}^2 $;
(\emph{ii}) setting everything as real, i.e.
$T=\Bar{T}=\hat{T}$, $W = \Bar{W}$, and $\check{T}=0$. Therefore we have
\begin{equation}
\begin{split}
    S^{3D}_{CF}~=&~ \int\,d^3x\, T ~+~ 2\,\int\,d^3x\,d^2\theta\,\Psi(T)\,W^2 ~~,
\end{split}
\label{equ:3D-CF}
\end{equation}
where $\Psi(T)$ takes the same form as \eqref{Psi} and (\ref{equ:3D-CF}) matches with the supersymmetric Born-Infeld action (\ref{equ:sBI-action}).

\section{Conventions in 3D, ${\cal N} = 1$ Superspace}
\label{3DSS}
In this appendix, we will briefly summarize our conventions and notations, which mostly follow from those of \cite{Gates:1983nr}.
In three dimensional spacetime, the Lorentz group is SL(2,R) and the corresponding fundamental representation acts on a real (Majorana) two-component spinor $\psi^{\alpha} = (\psi^+,\psi^-)$. Vector indices are denoted as $\underline{a} = 0, 1, 2$.

We choose Gamma matrices as
\begin{equation}
\begin{split}
(\gamma^0)_{\alpha}^{\ \beta} ~=&~ (i\sigma^2)_{\alpha}^{\ \beta} ~~~,\\
(\gamma^1)_{\alpha}^{\ \beta} ~=&~ (\sigma^3)_{\alpha}^{\ \beta} ~~~,\\
(\gamma^2)_{\alpha}^{\ \beta} ~=&~ (\sigma^1)_{\alpha}^{\ \beta} ~~~,\\
\end{split}
\end{equation}
which satisfy the Clifford algebra:
\begin{equation}
\{ \gamma^{\underline{a}}, \gamma^{\underline{b}} \} ~=~ 2 \eta^{\underline{a}\underline{b}} \mathbb{I} ~~~,
\end{equation}
where the Minkowski metric is
\begin{equation}
\eta_{\underline{a}\underline{b}} ~=~ \eta^{\underline{a}\underline{b}} ~=~ \begin{pmatrix}
-1 & 0 & 0 \\
0 & 1 & 0 \\
0 & 0 & 1
\end{pmatrix} ~~~.
\end{equation}
The gamma matrix has the following trace identity,
\begin{equation}
(\gamma^{\underline{a}})_{\alpha}^{\ \beta}(\gamma_{\underline{b}})_{\beta}^{\ \alpha} ~=~ 2\delta_{\underline{b}}^{\ \underline{a}} ~~~.
\end{equation}

We use the spinor metric to raise and lower spinor indices:
\begin{equation}
\begin{split}
&\psi_{\alpha} ~=~ \psi^{\beta}C_{\beta\alpha} ~~~,\\
&\psi^{\alpha} ~=~ C^{\alpha\beta}\psi_{\beta} ~~~,\\
\end{split}
\label{equ:raise-lower-spinior}
\end{equation}
where the definition of the spinor metric is
\begin{equation}
C_{\alpha\beta} ~=~ -C_{\beta\alpha} ~=~ -C^{\alpha\beta} ~=~\begin{pmatrix}
0 & -i \\
i & 0
\end{pmatrix}~~~.
\label{equ:spinormetric}
\end{equation}
From (\ref{equ:raise-lower-spinior}) and (\ref{equ:spinormetric}), we have the following identities
\begin{align}
	&C_{\alpha\beta}C^{\gamma\delta} ~=~ \delta_{[\alpha}^{\ \gamma}\delta_{\beta]}^{\ \delta} ~~~,\\
	&C_{\alpha\beta}C^{\alpha\delta} ~=~ \delta_{\beta}^{\ \delta} ~~~,\\
	&\psi^2 ~=~ \frac{1}{2}\psi^{\alpha}\psi_{\alpha} ~=~ i\psi^+\psi^- ~~~,\\
	&\psi^{\alpha}\psi_{\alpha} ~=~ -\psi_{\alpha}\psi^{\alpha} ~~~.
\end{align}

By using the spinor metric, we know that the gamma matrices are symmetric, namely,
\begin{equation}
\begin{split}
&(\gamma_{\underline{a}})_{\alpha\beta} ~=~ (\gamma_{\underline{a}})_{\beta\alpha} ~~~,\\
&(\gamma_{\underline{a}})^{\alpha\beta} ~=~ (\gamma_{\underline{a}})^{\beta\alpha} ~~~.
\end{split}
\end{equation}

Below, we list some useful identities of gamma matrices.
\begin{align}
&A_{[\alpha}B_{\beta]} ~=~ -C_{\alpha\beta} A^{\gamma}B_{\gamma}~~~,\\
&\gamma^{\underline{a}}\gamma_{\underline{a}} ~=~ 3\mathbb{I}~~~,\\
&\gamma_{\underline{a}}\gamma_{\underline{b}} ~=~ -\epsilon_{\underline{a}\underline{b}\underline{c}}\gamma^{\underline{c}} ~+~ \eta_{\underline{a}\underline{b}}\mathbb{I}~~~,\\
&\gamma^{\underline{b}}\gamma_{\underline{a}}\gamma_{\underline{b}} ~=~ -\gamma_{\underline{a}}~~~,\\
&(\gamma^{\underline{a}})_{\alpha\beta}(\gamma_{\underline{a}})^{\gamma\delta} ~=~ -\frac{3}{2}\delta_{\alpha}^{\ \gamma}\delta_{\beta}^{\ \delta} ~-~ \frac{1}{2}(\gamma^{\underline{a}})_{\alpha}^{\ \gamma}(\gamma_{\underline{a}})_{\beta}^{\ \delta} ~~~,\\
&(\gamma^{\underline{a}})_{\alpha\beta}(\gamma_{\underline{a}})^{\gamma\delta} ~=~ -\delta_{(\alpha}^{\ \gamma}\delta_{\beta)}^{\ \ \delta} ~=~ -(\gamma^{\underline{a}})_{(\alpha}^{\ \gamma}(\gamma_{\underline{a}})_{\beta)}^{\ \ \delta} ~~~,
\end{align}
where we define $\epsilon^{012} = 1$.

In the 3D, ${\cal N}=1$ superspace, the superspace coordinate is labeled by $z^A ~=~ (\,x^{\a\b},\,\theta^{\a}\,)$. They satisfy the hermiticity condition $(z^A)^\dagger = z^A$. Define derivatives as
\begin{equation}
    \begin{split}
        \pa_{\a\b}\,x^{\g\d}~\equiv&~ [\,\pa_{\a\b}\,,\, x^{\g\d}\,] ~=~ \fracm12\,\d_{(\a}^{\g}\,\d_{\b)}^{\d}\\
        \pa_{\a}\,\theta^{\b} ~\equiv&~ \{\,\pa_{\a}\,,\,\theta^{\b}\,\}~=~\d_{\a}^{\b}
    \end{split}
\end{equation}
implying that
\begin{equation}
    [\,\pa_{\a\b}\,]^{\dagger} ~=~ -\,\pa_{\a\b} ~~,~~[\,\pa_{\a}\,]^{\dagger} ~=~ \pa_{\a} ~~,~~[\,\pa^{A}\,]^{\dagger} ~=~ -\,\pa^{A} ~~.
    \label{equ:hermitian}
\end{equation}

The superspace covariant derivatives are defined as
${\rm D}_{A} ~=~ (\,\pa_{\a\b},\, {\rm D}_{\a}\,)$,
where
\begin{equation}
\begin{split}
    &\pa_{\a\b}~=~i\,(\g^{\un a})_{\a\b}\,\pa_{\un a}~~, \\
    &{\rm D}_{\a} ~=~ \pa_{\a} ~+~ i\,\theta^{\b}\,\pa_{\a\b}~~.\\
\end{split}
\end{equation}
They satisfy the algebra
\begin{equation}
    \begin{split}
        &\{\,{\rm D}_{\a}\,,\,{\rm D}_{\b}\,\} ~=~ 2i\,\pa_{\a\b}~~,\\
    &[\,\pa_{\a\b}\,,\,{\rm D}_{\g}\, ]~=~ 0~~.
    \end{split}
\end{equation}

Finally, we list some identities of covariant derivatives, which are useful in the calculations we have encountered throughout this paper.
\begin{align}
    &\pa^{\a\g}\,\pa_{\b\g}~=~ \d_{\b}^{\a}\,\Box~~,\\
    &{\rm D}_{\a}\,{\rm D}_{\b}~=~i\,\pa_{\a\b}~-~C_{\a\b}\,{\rm D}^2~~,\\
    &{\rm D}^{2}\,{\rm D}_{\a}~=~ - {\rm D}_{\a}\,{\rm D}^{2}~=~ i\,\pa_{\a\b}\,{\rm D}^{\b}~~,\\
    & {\rm D}^{\b}\,{\rm D}_{\a}\,{\rm D}_{\b}~=~0~~,\\
    & ({\rm D}^{2})^2 ~=~ \Box~~,
\end{align}
where
\begin{equation}
   \begin{split}
       \Box~=&~ \fracm12\,\pa^{\a\b}\,\pa_{\a\b}
       ~=~ \pa^{\un a}\,\pa_{\un a}~~,\\
       {\rm D}^2 ~=&~ \fracm12\,{\rm D}^{\a}\,{\rm D}_{\a}~~.
   \end{split}
\end{equation}

% Create the reference section using BibTeX:
%\begin{multicols}{2}
%{\small
%\bibliographystyle{hephys}
%\bibliography{references}}
%\end{multicols}

\bibliographystyle{hephys}
\bibliography{references}

\end{document}